\documentclass[journal,comsoc]{IEEEtran}
\usepackage[T1]{fontenc}
\ifCLASSINFOpdf
   \usepackage[pdftex]{graphicx}
\else
\fi
\usepackage{amsmath}
\usepackage{algorithmic}
\usepackage{array}
\usepackage{color, verbatim, multirow}
\usepackage[linesnumbered,ruled,vlined]{algorithm2e}
\usepackage[normalem]{ulem}
\newtheorem{theorem}{Theorem}
\newtheorem{lemma}{Lemma}
\newtheorem{define}{Definition}

\newtheorem{proposition}{Proposition}
\newtheorem{corollary}{Corollary}

\begin{document}

\title{Survivable Cloud Network Design Against Multiple Failures Through Protecting Spanning Trees}
\author{
        Zhili~Zhou,~\IEEEmembership{Member,~IEEE,}
        Tachun~Lin,~\IEEEmembership{Member,~IEEE,}
        and~Krishnaiyan~Thulasiraman,~\IEEEmembership{Fellow,~IEEE}
\thanks{Zhili~Zhou is with the United Airlines.}
\thanks{Tachun~Lin is with the Department of Computer Science and Information Systems, Bradley University, Peoria, IL 61625, USA  (e-mail: djlin@bradley.edu.)}
\thanks{Krishnaiyan~Thulasiraman is with the School of Computer Science, University of Oklahoma, Norman, OK 73019, USA.}}
\maketitle

\begin{abstract}
Survivable design of cross-layer networks, such as the cloud computing infrastructure, lies in its resource deployment and allocation and  mapping of the logical (virtual datacenter/IP) network into the physical infrastructure (cloud backbone/WDM) such that link or node failure(s) in the physical infrastructure would not result in cascading failures in the logical network.
Most of the prior approaches for survivable cross-layer network design aim at single-link failure scenario, which are not applicable to the more challenging multi-failure scenarios. Also, as many of these approaches use the cross-layer cut concept, enumeration of all cuts in the network is required and thus introducing exponential number of constraints. To overcome these difficulties, we investigate in this paper survivable mapping approaches against multiple physical link failures and its special case, Shared Risk Link Group (SRLG) failure. We present the necessary and sufficient conditions based on both cross-layer spanning trees and cutsets to guarantee a survivable mapping when multiple physical link failures occur. Based on the necessary and sufficient conditions, we propose to  solve the problem through (1) mixed-integer linear programs which avoid enumerating all combinations of link failures, and (2) an algorithm which generates/adds logical spanning trees sequentially. Our simulation results show that the proposed approaches can produce survivable mappings effectively against both $k$- and SRLG-failures.

\end{abstract}

\section{Introduction}
\begin{figure}
\centering
\includegraphics[scale=1]{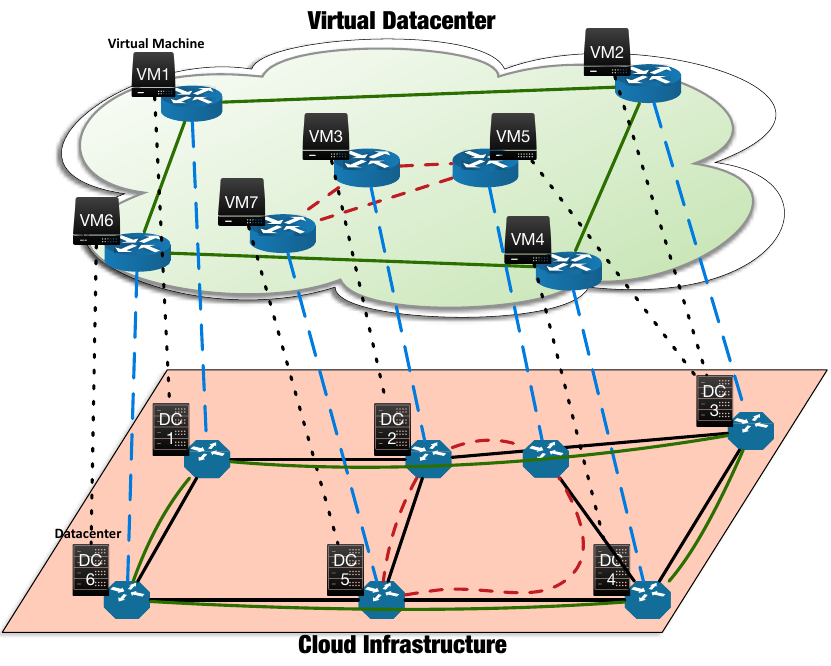}
\caption{Cloud computing infrastructure}
\label{fig:cloudNet}
\end{figure}

Cloud computing through which demands of geographically distributed individuals and enterprise are realized seamlessly over a common physical cloud infrastructure/backbone has attracted significant attention in recent years both in academia and industry. Cloud service providers offer their users an abstracted layer of computational (processors/memory) and communication resources considered as a virtual datacenter, and cloud services are then carried out through virtual-machine deployment and its mapping to the physical datacenters and their data connections in the cloud infrastructure~\cite{develder2012optical}\cite{VirtualNetwork2009Chowdhury}. Figure \ref{fig:cloudNet} illustrates a cloud computing infrastructure, where the virtual datacenter (top-layer) consists of interconnected virtual machines is mapped onto the cloud infrastructure (bottom-layer) with datacenters and communication networks. As each cloud service provider usually supports multiple tenants, each virtual machine is allocated with a fraction of the computational capacity of the correspondingly mapped datacenters, and the links connecting virtual machines would also occupy a portion of the overall capacity of the fiber-optic connections at the cloud backbone. Hence, failure(s) in the physical infrastructure may disconnect both physical datacenters and virtual machines and decrease the computational and communication capacities of the cloud. This has motivated studies on the problems of resilient and survivable network design.

A virtual datacenter is claimed to be survivable if its services sustain when failure(s), such as link failures in optical networks~\cite{colman2014disaster} or power outage in datacenters~\cite{habib2012}, occur in the cloud infrastructure. In this paper, we use network survivability as a measure to identify if a network remains connected against instantaneous failures~\cite{ramanurthy03}{\cite{yu2010survivable}\cite{rahman2013svne}\cite{soualah2013pr}}. The approaches to guarantee virtual datacenter survivability can be achieved in different layers. Since the optical network is the major communication media in a cloud infrastructure connecting datacenters, the first approach is to protect the internet protocol (IP) traffic over optical networks {or virtual network traffic over physical substrate network through backup protection}, which ensures that the {fiber/phyiscal link} is protected by either a dedicated or shared backup {fiber/physical link}. {1+1, 1:1, and 1:$N$ protection~\cite{ietf-mpls-01} are the traditional but most commonly applied approaches. For the link-failure case, two-stage approaches with reactive re-routing are explored in~\cite{rahman2010survivable}\cite{LinZho14}, and network/service backup provisioning as proactive approaches are proposed in~\cite{guo2011shared}\cite{chen2010resilient}. As a node failure results in multiple link failures, the failure-dependent protection is introduced in~\cite{yu2010survivable}\cite{qiao2011novel} where backup nodes are utilized; and the failure-independent protection discussed in~\cite{liu2009robust}\cite{yeow2011designing}\cite{guo2014survivable}\cite{sun2015power}\cite{chowdhury2016protecting} targets to provide the same level of protection under a single node/link failure scenario with less substrate resources.}
{A closely related work on reliable cross-layer network design emphasizes controlling the failure probability of the entire network considering link failure probability. Probabilistic analysis~\cite{lee2011reliability} and scenario-based approaches~\cite{anastasopoulos2013stochastic}\cite{luo2014survivable}\cite{meixner2013disaster}\cite{dikbiyik2014minimizing}  are proposed for multiple link failures and disaster recovery.}

{Different from backup-provisioning and probability-based approaches mentioned above, our cross-layer survivable routing scheme maximally utilizes existing resources and protects the topology through routing.}
Due to its NP-completeness \cite{NPC}, early works are mostly concentrated on single physical link failure scenario. Kurant and Thiran~\cite{KurThi04} proposed a disjoint-path-based protection scheme to guarantee the sufficient condition for cross-layer network survivability, where logical cycles are mapped onto link-disjoint paths iteratively. Zhou et al.~\cite{ZhoLin12}\cite{ICC15} provided necessary and sufficient conditions for the existence of survivable design using the concept of protecting spanning trees, while Modiano and Narula-Tam~\cite{ModNar01} proposed a different approach based on cross-layer cutsets. To address the scenario of multiple physical link failures, Todimala and Ramamurthy~\cite{TodRam07} studied survivable mappings against a single SRLG failure. {The approach proposed in~\cite{TodRam07} is also based on cross-layer cutsets introduced in~\cite{ModNar01}.} Here SRLG means a group of network links routed through the same fiber, thus the fiber cut causes simultaneous failures to this group of links.
{Jaumard et al.~\cite{jaumard2012path} and Parandehgheibi et al.~\cite{parandehgheibi2014survivable} proposed the use of minimal path sets for survivable routing after single link failure.} Xi et al.~\cite{SRLG10Xi} considered rerouting as a restoration scheme to recover SRLG failures in IP-over-WDM networks.
Similar concepts have also been applied to the survivable cloud network mapping problem{~\cite{rahman2013svne}. For instance, an integrated approach with content placement/replica and routing was introduced in~\cite{habib2012}\cite{pourvali2016progressive}, anycast routing presented in~\cite{bui2013anycast}\cite{wu2013multicast}, and multi-path routing schemes discussed in~\cite{khan2015multi}\cite{khan2015simple}}.

Multiple physical link failures bring in more challenges for cloud network mapping because (1) none of the survivable mapping designs for single failure would work if a link mapping is routed through multiple failed physical links; (2) compared with a single physical link failure scenario which has in total $|E_P|$ possible single physical link failures, there exist $ C^{|E_P|}_k$ number of $k$-edge combinations under the $k$-failure scenario, which all need to be considered; (3) intuitively, to capture full information for $k$-link failures, where $k\in{1,\ldots,|E_P|}$, all possible $k$-link combinations should be enumerated, which has an exponential number of combinations $(2^{|E_P|}-1)$; and last but not least, even though there are only $|R_E|$ number of failure combinations in the SRLG scenario, each element in $R_E$ may include different number of links. Thus, it is important to explore a solution approach for above problems without enumerating all possible combinations, which is the key question we would like to answer.

In this paper, we study the survivable cloud network mapping problem against the generalized $k$ physical link failures and its special case, the SRLG failure. We propose the necessary and sufficient conditions for survivable cloud network mapping against multiple physical link failures and design the exact solution approach through mixed-integer linear program (MILP) formulations. The contributions of this paper are as follows. We first study the survivable network mapping problem against multiple-link failure scenarios and prove the necessary and sufficient conditions for $k$- and SRLG-survivability under a cross-layer network setting. {Second, different from the approaches in~\cite{ModNar01}\cite{TodRam07}}, our proposed exact solution approaches do not enumerate all possible combinations of failed physical link sets, which greatly reduces the computation time. In addition to the mathematical formulation approach, we also design a heuristic which can sequentially generate and select logical spanning trees as well as their corresponding tree branch mappings.

The rest of paper is organized as follows. Section \ref{sec:ProblemDescription} provides the network setting in a cloud infrastructure and definitions of $k$- and SRLG-survivability. Necessary and sufficient conditions {for cross-layer network survivability against} $k$- and SRLG- failures are presented in Section~\ref{sec:nsCondition}. From the necessary and sufficient conditions we propose in Section~\ref{sec:Formulation} the MILP formulations realizing the protecting spanning tree concept for both $k$- and SRLG- failures. In Section \ref{sec:solutionApp}, we present an SRLG-survivable algorithm which sequentially selects protecting spanning trees and generates their corresponding mappings. We also demonstrate its extension to $k$-survivability. Finally, simulation results for both the MILP model and proposed heuristic algorithm are reported in Section \ref{sec:computation}.

\section{Network Setting and Survivability}\label{sec:ProblemDescription}
\begin{table}[!ht]
\begin{center}
\begin{tabular}{p{2cm}|p{6cm}}
\hline
 \rule{0pt}{11pt}Notation & Representation\\
\hline
 \rule{0pt}{11pt}$G_P$ & The physical cloud infrastructure, $G_P=(V_P, E_P)$  with $V_P$ and $E_P$ as its node and edge sets\\
 \rule{0pt}{11pt}$G_{S}$ & The virtual datacenter network, $G_{S}= (V_{S}, E_{S})$ with $V_{S}$ and $E_{S}$ as its node and edge sets\\
 \rule{0pt}{11pt}$(i,j)$ & The physical link, $i, j\in V_P, (i,j)\in E_P$\\
 \rule{0pt}{11pt}$(s,t)$ & The virtual link, $s, t\in V_S, (s,t)\in E_S$\\
 \rule{0pt}{11pt}$R_{E}$& Shared risk link groups (SRLGs), whose element $r= \{(e^{r}_{1}, e^{r}_{2}, \ldots, e^{r}_{k}): e^{r}_{\ell} \in E_P\}, r \in R_{E}$ represents an SRLG\\
 \rule{0pt}{11pt}$\mathcal{T}$& Spanning tree set, $\mathcal{T}\subset G_{S}$, whose element $\tau\in\mathcal{T}$ is a spanning tree\\
 \rule{0pt}{11pt}$p_{st}$ & The routing of $(s,t)\in E_S$ in $G_P$, where $p_{st}\subset E_P$\\
 \rule{0pt}{11pt}$\Lambda(i,j)$& A set of links in $G_S$ which are routed through physical link $(i,j)$, i.e., $\{(s,t): (i,j)\in p_{st}, (s,t)\in E_{S}\}$ \\
\\ \hline
\rule{0pt}{11pt} Variables & Representation\\
\hline \\
\rule{0pt}{12pt} $y^{st}_{ij}$&binary variable which indicates whether $p_{st}$ is routed through $(i,j)$ or not, where $(i,j)\in E_P $; if yes, $y^{st}_{ij}=1$, otherwise $y^{st}_{ij}=0$\\
\rule{0pt}{12pt} $y^{st}_{i_1j_1,\cdots,i_kj_k}$ & binary variable which indicates whether $p_{st}$ is routed through any links in $(i_1,j_1)$,$\cdots$,$(i_k,j_k)$; if yes, $y^{st}_{i_1j_1,\cdots,i_kj_k}=1$, otherwise $y^{st}_{i_1j_1,\cdots,i_kj_k}=0$\\
\rule{0pt}{12pt} $\mu_{i_1j_1, i_2j_2, \ldots, i_kj_k}^{st}$&binary variable which indicates if $p_{st}$ is \textbf{not} routed through any of the links in $(i_1,j_1)$,$\cdots$,$(i_k,j_k)$; if yes, $\mu_{i_1j_1, i_2j_2, \ldots, i_kj_k}^{st}= 1$, otherwise $\mu_{i_1j_1, i_2j_2, \ldots, i_kj_k}^{st}=0$
\\ \hline
\end{tabular}
\caption{Notations}\label{tab:notation}
\end{center}
\end{table}

\begin{figure}[!ht]
\begin{center}
\includegraphics[scale=1]{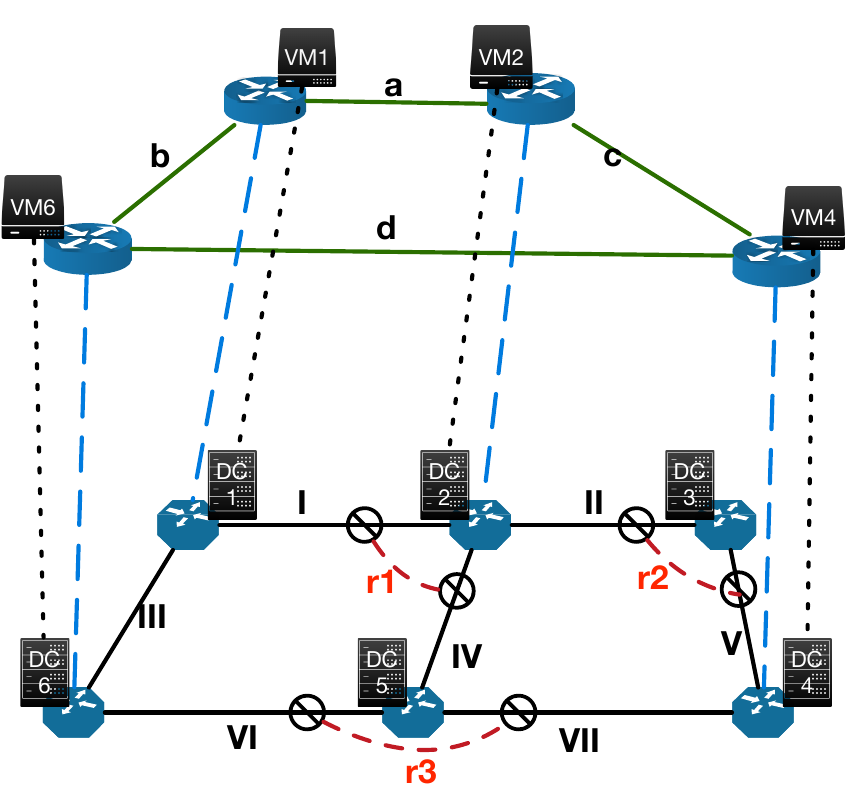}
\end{center}
\caption{Survivable cloud network mapping with SRLG failure}
\label{fig:srlgFigure}
\end{figure}

Notations utilized in this paper are presented in Table~\ref{tab:notation}. Let $G_{S} = (V_{S}, E_{S}), G_P = (V_P, E_P)$ denote the virtual datacenter and physical cloud infrastructure, respectively, where $V_S, V_P$ represent the node sets and $E_S, E_P$ represent the link sets in their corresponding network. Each $s\in V_{S}$ is mapped onto a node $u\in V_P$, while requests between nodes $s$ and $t$, represented as a link $(s,t)\in E_{S}$, are realized through a route (denoted as $p_{st}$) in $G_P$ connecting $s$ and $t$'s corresponding nodes in $V_P$. Here we use \textbf{\textit{cloud network mapping}} to represent both node and link mappings of the virtual datacenter. Let $R_E$ denote {an} SRLG set containing all SRLG failure scenarios, where each element $r\in R_E$ represents a single SRLG. Let $k(r)$ represent the number of physical links in SRLG $r$; i.e., $r=\{(i_1, j_1), (i_2, j_2), \cdots, (i_{k(r)}, j_{k(r)})\}$.
\begin{define}\label{def:srlgSurv}
A cloud {network} mapping is \textbf{\textit{SRLG-survivable}} if {the virtual datacenter} remains connected after any SRLG failure in $R_E$.
\end{define}
\begin{define}\label{def:kSurv}
A cloud {network} mapping is \textbf{\textit{k-survivable}} if the {virtual datacenter} remains connected after arbitrary $k$ physical link failures.
\end{define}
For the virtual datacenter to survive $k$- and SRLG-failures, at least a spanning tree $\tau$ should be available in its residual network after failures. We use Fig.~\ref{fig:srlgFigure} to {illustrate} an SRLG-survivable cloud network mapping.

Given the SRLG failure set $R_E=\{r_1, r_2, r_3\}$ with {$r_1=\{I, IV\}$, $r_2=\{II, V\}$}, and {$r_{3}=\{VI, VII\}$}. Each virtual machine in the virtual datacenter is mapped onto the physical datacenter with the same index, and the virtual links $\{a, b, c, d\}$ connecting virtual machines mapped onto physical paths {$p_{a}=\{I\}$, {$p_{b}=\{III\}$}, $p_{c}=\{II, V\}$, and $p_{d}=\{VI, VII\}$}. After {$r_1$'s} failure, a spanning tree {$\tau_1=\{b, c, d\}$} is available in the virtual datacenter, which makes it survivable. With the existence of two other spanning trees {$\tau_2=\{a, b, d\}$} and {$\tau_3=\{a, b, c\}$}, the virtual datacenter also remains connected after SRLG $r_2$ or $r_3$ failures. Hence, with the given mapping, the virtual datacenter is survivable after any SRLG failure in $R_E$, and we call spanning trees $\tau_1, \tau_2, \tau_3$ the \textit{\textbf{protecting spanning trees}}.

\section{Necessary and Sufficient Conditions for $k$- and SRLG-Survivability}\label{sec:nsCondition}
{We present in this section the necessary and sufficient conditions based on the spanning tree concept for both {$k$}- and SRLG-survivability. We also include in Appendix the cutset-based necessary and sufficient conditions.}

\subsection{Survivability Based on Protecting Spanning Trees}\label{subsec:proTree}
For any $(i,j)\in E_P$, we let $\Lambda(i,j)$ denote the set of link{s} $(s,t)\in E_S$ whose {routes pass} through $(i,j)$, i.e., $\Lambda(i,j)=\{(s,t): (i,j)\in p_{st}, (s,t)\in E_S\}$. Let {$\tau$ {be} a spanning tree in the virtual datacenter where $\tau\subset G_S$. We now derive the spanning-tree based necessary and sufficient conditions for $k$- and SRLG-survivability as follows.}

\begin{lemma}\label{lemma:1tree}
A cloud network mapping is 1-survivable if and only if after the failure of each $(i,j)\in E_P$, there exists at least a spanning tree $\tau \subset G_S$, such that
\begin{align}
\tau \cap \Lambda(i,j)=\emptyset. \nonumber
\end{align}
\end{lemma}
This lemma originated from the fact that a mapping is survivable if and only at least a spanning tree is embedded in the virtual datacenter after any single physical link failure (see~\cite{LinThu12}).

Extending this {lemma}, we get the following necessary and sufficient conditions for $k$-survivability.
\begin{theorem}\label{thm:ksurv}
A cloud network mapping is $k$-survivable if and only if {there exists at least a spanning tree $\tau\subset G_S$ after} arbitrary $k$ physical link {failures}, such that
\begin{align}
\tau \bigcap \left(\bigcup_{\beta=1}^{k}\Lambda(i_\beta, j_\beta)\right)=\emptyset,\nonumber
\end{align}
where $(i_1, j_1)\neq (i_2, j_2)\neq \cdots\neq (i_k, j_k)$ and $(i_1, j_1), (i_2,j_2), \cdots, (i_k,j_k)\in E_P$.\label{treeIneq_k}
\end{theorem}
\begin{IEEEproof}
{Necessary condition: {if} after any {$k$-link failures,} at least a spanning tree structure is embedded in the residual virtual datacenter, then the virtual datacenter remains connected. Hence, the cloud network mapping is $k$-survivable. \\
Sufficient condition: If a cloud network mapping is $k$-survivable, the virtual datacenter remains connected after any $k$-link failures. Hence, at least a spanning tree structure exists in the residual virtual datacenter.}
\end{IEEEproof}

\begin{corollary}\label{co:treeTheorem}
A cloud network mapping is SRLG-failure survivable if and only if after any $r\subset R_E$ failure, there exists at least a spanning tree $\tau\subset G_S$, such that
\begin{align}
\tau\bigcap \left(\bigcup_{(i^{r}_{k(r)}, j^{r}_{k(r)})\in r}\Lambda(i^{r}_{k(r)},j^{r}_{k(r)})\right) = \emptyset.\label{treeIneql}
\end{align}
\end{corollary}
{The above conclusion is a direct extension of Theorem \ref{thm:ksurv} with the fact that {the} SRLG failure is {a} special case of the generalized $k$-failure scenario.}

We may now introduce our solution approaches based on the necessary and sufficient conditions presented in this section.

\section{Solution Approach I: MILP Formulations}
\label{sec:Formulation}
In this section, we first present a mathematical programming formulation for the survivable cloud network mapping problem when arbitrary $k$-link failures occur in the physical cloud infrastructure. We then present a formulation for the case of SRLG failures.

We first introduce the variables used in the MILP formulations. Let $y^{st}_{ij}$ be {a} variable which represents whether the {mapping} of $(s,t)\in E_S$ is routed through $(i,j)\in E_P ${;} if yes, $y^{st}_{ij}=1${,} otherwise $y^{st}_{ij}=0${. Let} $y^{st}_{i_1j_1,\cdots,i_kj_k}$ be a variable which equals 1 if $(s,t)$'s corresponding route in $G_P$ passes through one or more of the links $(i_1,j_1),\cdots, (i_k,j_k)$, where $(s,t)\in E_S$, $(i_1,j_1)\neq (i_2,j_2)\neq \cdots\neq (i_k,j_k)$ and $(i_\beta, j_\beta)\in E_P$ with $1\leq \beta\leq k$; otherwise, this variable {equals} 0.
We let $\mu^{st}_{ij}$ be a variable that {equals} 0 if the mapping of link $(s,t)\in E_S$ is routed through $(i,j)\in E_P$; otherwise, $0 < \mu^{st}_{ij} \leq 1$. {Let} $\mu_{i_1j_1, i_2j_2, \ldots, i_kj_k}^{st}$ be a variable which equals 0 if the link $(s,t)\in E_S$ is disconnected after the failure of one or more of the links $(i_1, j_1), \ldots, (i_k, j_k)$, where $(i_1,j_1)\neq (i_2,j_2)\neq \ldots\neq (i_k,j_k)$ and $(i_\beta, j_\beta)\in E_P$ with $1\leq \beta\leq k$; otherwise, $0< \mu_{i_1j_1, i_2j_2, \ldots, i_kj_k}^{st}\leq 1$.

To express the condition in Theorem \ref{thm:ksurv} in terms of variable $y$, we propose the following lemmas which build the connection between links in the virtual datacenter and those in the cloud infrastructure.

\begin{lemma}\label{lemma:2FailureMset}
Given $(i_1, j_1), (i_2,j_2)\in E_P, (i_1, j_1)\neq (i_2,j_2)$ and $(s,t)\in E_S$. $(s,t)$ has the property $(s,t) \in \Lambda(i_1, j_1)\cup \Lambda(i_2, j_2)$ if and only if $y^{st}_{i_1j_1,i_2j_2}=1$, which {can be formulated by}
\begin{align}
y^{st}_{i_1j_1,i_2j_2} &\geq y^{st}_{i_1j_1} + y^{st}_{j_1i_1},\label{fm:2FailMseta}\\
y^{st}_{i_1j_1,i_2j_2} &\geq y^{st}_{i_2j_2}+y^{st}_{j_2i_2}, \label{fm:2FailMsetb}\\
y^{st}_{i_1j_1,i_2j_2} &\leq  y^{st}_{i_1j_1}+y^{st}_{j_1i_1}+ y^{st}_{i_2j_2}+y^{st}_{j_2i_2}. \label{fm:2FailMsetc}
\end{align}
\end{lemma}
Proof of lemma~\ref{lemma:2FailureMset} is given in Appendix C.

Extending Lemma \ref{lemma:2FailureMset}, the following lemma shows the property between a single virtual datacenter link and $k$ links in the cloud infrastructure.
\begin{lemma}\label{lemma:kFailureMset}
Given $(i_1, j_1),\ldots, (i_k,j_k)\in E_P, (i_1, j_1)\neq (i_2,j_2)\neq \ldots \neq (i_k, j_k)$ and $(s,t)\in E_S$. $(s,t)$ has the property $(s,t)\in \bigcup_{\beta=1}^{k}\Lambda(i_\beta,j_\beta)$ if and only if $y^{st}_{i_1j_1,\cdots, i_kj_k}=1${,} which {can be formulated by}
\begin{align}
&y^{st}_{i_1j_1, \ldots, i_k j_k} \geq y^{st}_{i_1j_1} + y^{st}_{j_1i_1},\label{beta_fail_1}\\
&\ldots\ldots&\;\nonumber\\
&y^{st}_{i_1j_1, \ldots, i_k j_k} \geq y^{st}_{i_k, j_k}+y^{st}_{j_ki_k},\label{beta_fail_beta}\\
&y^{st}_{i_1j_1, \ldots, i_k j_k} \leq
\sum_{\beta=1}^{k}(y^{st}_{i_\beta j_\beta}+y^{st}_{j_\beta i_\beta}). \label{beta_fail_all}
\end{align}
\end{lemma}
Proof of this lemma is presented in Appendix C.

With Lemma~\ref{lemma:kFailureMset}, we may now derive the formulations for $k$- and SRLG-survivable cloud network mapping.
\begin{proposition}\label{prop:formulation}
The {relationship} between $\mu_{i_1j_1, i_2j_2, \ldots, i_kj_k}^{st}$ and $y^{st}_{i_1j_1},\ldots, y^{st}_{i_kj_k}$ {can be} captured by the following constraints:
\begin{align}
&\mu_{i_1j_1, i_2j_2, \ldots, i_kj_k}^{st} \leq 1 - (y^{st}_{i_1j_1} +
y^{st}_{j_1i_1}), &\label{ksurvReCapa_1}\\
&\mu_{i_1j_1, i_2j_2, \ldots, i_kj_k}^{st} \leq 1 - (y^{st}_{i_2j_2} +
y^{st}_{j_2i_2}), &\label{ksurvReCapa_2}\\
&\ldots\ldots &\nonumber \\
&\mu_{i_1j_1, i_2j_2, \ldots, i_kj_k}^{st} \leq 1 - (y^{st}_{i_kj_k} +
y^{st}_{j_ki_k}), &\label{ksurvReCapa_k}\\
&(s,t)\in E_{S}, (i_1, j_1)\neq (i_2, j_2)\neq \cdots\neq (i_k, j_k) \nonumber\\
&\text{ and } (i_\beta, j_\beta)\in E_P, y^{st}_{i_{\beta} j_{\beta}}\in \{0,1\}, 1\leq \beta \leq k. &\label{bounds}
\end{align}
\end{proposition}
With {variables} $y$ and $\mu$, the feasible region of $y^{st}_{ij}$, $Y =\{y^{st}_{ij}: \text{{Constraints} } (\ref{eq:lightpath1}) \text{ and } (\ref{eq:feasible_y}), \text{ with } i\in V_{P},\}$, determines a cloud network {mapping}, where
\begin{align}
&\sum_{(i,j)\in E_P}y^{st}_{ij}-\sum_{(j,i)\in E_P}y^{st}_{ji}=
\left\{\begin{matrix}
1, &\,\mbox{if } i=s,\\
-1, &\,\mbox{if } i=t,\\
0, &\,\mbox{if } i\neq \{s, t\},
&\end{matrix}\right.\label{eq:lightpath1}\\
& y^{st}_{ij}\in \{0,1\}, (s,t)\in E_S, (i,j)\in E_P.\label{eq:feasible_y}
\end{align}

\begin{theorem}\label{thm:k-surv}
A cloud network mapping is $k$-survivable {against} arbitrary $k$ physical link failures{,} $k\geq 1$, if and only if the following {conditions} hold.
\begin{align}
&\mu_{i_1j_1,i_2j_2,\cdots,i_{k} j_{k}}^{st} \leq 1 - (y^{st}_{i_{\beta}j_{\beta}}+y^{st}_{j_{\beta}i_{\beta}}),\quad 1\leq \beta\leq k \label{fm:Beta_1}\\
&\sum_{(s,t)\in E_{S}}\mu_{i_{1}j_{1},\ldots, i_{k}j_{k}}^{st} - \sum_{{(t,s)}\in E_{S}}\mu_{i_{1}j_{1},\ldots, i_{k}j_{k}}^{ts} \nonumber\\
= &\left\{\begin{array}{c}
            -1,\quad\quad\quad\quad\quad\quad\quad\quad\quad\quad s= v_0, v_0\in V_{S} \\
            \frac{1}{|V_{S}| -1},  \quad\quad\quad\quad\quad\quad\quad\quad\;\; s \neq v_0, v_0\in V_{S}
          \end{array} \right. \label{fm:treeBalance}\\
& 0\leq \mu_{i_1j_1, i_2j_2, \ldots, i_{k}j_{k}}^{st} \leq 1, \quad\quad\quad\quad\;\; (s,t)\in E_{S} \label{fm:treeFeasible}
\end{align}
\end{theorem}
Constraint (\ref{fm:treeBalance}) captures the connectivity requirements (existence of a spanning tree).

Hence, the exact solution approach {in} MILP for the $k$-survivable cloud network mapping {against} multiple physical link failures is as follows:
\begin{align}
&\min \sum_{(s,t)\in E_{S}}\sum_{(i,j)\in E_P}y^{st}_{ij}\nonumber\\
s.t.\;& \mbox{ Constraints (\ref{bounds}) to (\ref{fm:treeFeasible}).}\nonumber
\end{align}

Based on Corollary \ref{co:treeTheorem} and Theorem \ref{thm:k-surv}, we obtain the following theorem which is a special case of Theorem~\ref{thm:k-surv} applicable to the SRLG failure case.

\begin{theorem}\label{Ksurvivable}
A cloud network mapping is survivable after SRLG failures, if and only if the following conditions are satisfied{. For} any $r\in R_E$,
\begin{align}
&\mu_{i^{r}_{1}j^{r}_{1}, i^{r}_{2}j^{r}_{2}, \ldots, i^{r}_{k(r)}j^{r}_{k(r)}}^{st} \leq 1 - (y^{st}_{i^{r}_{1}j^{r}_{1}} +
y^{st}_{j^{r}_{1}i^{r}_{1}}) \label{ksurvReCapa11} \\
&\mu_{i^{r}_{1}j^{r}_{1}, i^{r}_{2}j^{r}_{2}, \ldots, i^{r}_{k(r)}j^{r}_{k(r)}}^{st} \leq 1 - (y^{st}_{i^{r}_{2}j^{r}_{2}} +
y^{st}_{j^{r}_{2}i^{r}_{2}}) \label{ksurvReCapa21} \\
&\ldots\ldots \nonumber \\
&\mu_{i^{r}_{1}j^{r}_{1}, i^{r}_{2}j^{r}_{2}, \ldots, i^{r}_{k(r)}j^{r}_{k(r)}}^{st} \leq 1 - (y^{st}_{i^{r}_{k(r)}j^{r}_{k(r)}} +
y^{st}_{j^{r}_{k(r)}i^{r}_{k(r)}}) \label{ksurvReCapa31}
\end{align}
\begin{align}
&\sum_{(s,t)\in E_{S}}\mu_{i^{r}_{1}j^{r}_{1},\ldots, i^{r}_{k(r)}j^{r}_{k(r)}}^{st} - \sum_{{(t,s)}\in E_{S}}\mu_{i^{r}_{1}j^{r}_{1},\ldots, i^{r}_{k(r)}j^{r}_{k(r)}}^{ts} \nonumber\\
= &\left\{\begin{array}{c}
            -1,\quad\quad\quad\quad\quad\quad\quad\quad\quad\;\,  s= v_0, v_0\in V_{S} \\
            \frac{1}{|V_{S}| -1},  \quad\quad\quad\quad\quad\quad\quad\;\;\; s \neq v_0, v_0\in V_{S}
          \end{array} \right. \label{treeBalance}\\
& 0\leq \mu_{i^{r}_1j^{r}_1, i^{r}_2j^{r}_2, \ldots, i^{r}_{k(r)}j^{r}_{k(r)}}^{st} \leq 1, \quad\quad (s,t)\in E_{S} \label{rrange}
\end{align}
where $v_0$ is a selected root node in the virtual datacenter.
\end{theorem}
Now we have the following MILP formulation for survivable cloud network mapping under SRLG failures which also minimizes the physical link utilization.
\begin{align}
& \min \;\sum_{(s,t) \in E_{S}}\sum_{(i,j)\in E_P} y^{st}_{ij} \nonumber\\
s.t.\; &\mbox{Constraints (\ref{eq:lightpath1}), (\ref{eq:feasible_y}), (\ref{ksurvReCapa11}) to (\ref{rrange})}\nonumber
\end{align}

We wish to note that {the cutset-based} formulations are provided in Appendix A. The reasons why we chose the tree-based formulation presented above instead of the cutset-based one are that (1) all variables $\mu$ are fractional and (2) the total number of binary variables is $|E_P||E_S|$. It can be observed that the cutset-based formulations are (1) with all binary variables and (2) required to enumerate all cutsets{,} which significantly increase the computational complexity of the MILP formulation.

\section{Solution Approach II: Heuristic Algorithm}\label{sec:solutionApp}
In this section, we first present the concept of optimal protecting spanning tree collection and how it works for {$k$- and SRLG-survivability}. Notations {used} in the following discussions are as follows. Let $\mathcal{P}$ be the cloud {network mapping} with $\mathcal{P}=\{p_{st}, (s,t)\in E_S\}$. Let $R^{k}_{F}$ {indicate} the failure set for all $k$-link {failures}, and $r$ be an element in $R^{k}_{F}$. For spanning tree $\tau \subset G_S$, we let $E_P(\tau)=E_P\setminus \{\cup_{(s,t)\in \tau}p_{st}\}$ and $R^{k}_{F}(\tau)=\{r, r\in R^{k}_{F}, {r\subseteq} E_P\setminus \{\cup_{(s,t)\in \tau}p_{st}\}\}$ denote the {physical links and the SRLG failure sets protected by $\tau$}, respectively. We let $\mathcal{T}=\{\tau: (s,t)\in \tau \text{ with } p_{st}\}$ be the spanning tree collection in the virtual datacenter.

\subsection{Protecting Spanning Tree Collection and Failed Physical Link Set}\label{subsec:1mMapping}
For SRLG failures, links in an SRLG fail simultaneously, thus a set of these links is called a failure set. Given an SRLG set $R_E$, $|R_E|$ represents the total number of elements in $R_E$ and $C^{|E_P|}_{k}$ is the total number of combinations for arbitrary $k$-link failures.  In this section, we investigate the relation between protecting spanning trees and failure set.

\begin{corollary}\label{pro:srlgTreeSurv}
Given a cloud network and an SRLG set $R_E$. If a cloud network is SRLG-survivable, there exists at least one spanning tree $\tau \subseteq G_S$ such that $r\in E_P(\tau)$ with $r\in R_E$.
\end{corollary}
Corollary~\ref{pro:srlgTreeSurv} presents the concept that at least one spanning tree $\tau$ which remains connected after the failures of an SRLG $r\in R_E$, thus $\tau$ is said to protect all the physical links in $r$. We may derive the following theorem addressing the maximal number of spanning trees required to protect $R_E$.
\begin{theorem}\label{prop:numTreeSRLG}
Given a cloud network and an SRLG set $R_E$. The sufficient condition to guarantee an SRLG-survivable cloud network design is that there exists up to $|R_E|$ number of spanning trees, where each spanning tree's branch routings only travel through physical links in $E_P\setminus r, r\in R_E$ .
\end{theorem}
Theorem~\ref{prop:numTreeSRLG} provides an observation that if each SRLG $r\in R_E$ is uniquely protected by one spanning tree, then we need in total $|R_E|$ spanning trees to protect the whole $R_E$, which is the upper bound. In other words, under this special optimal condition, adding more spanning tree will not help further in terms of survivability. In practice, each selected spanning tree may protect not only one but more SRLGs. Therefore, we will evaluate in Section~\ref{sec:computation} the number of spanning trees required to protect multiple failures.

Following the same concept as in Theorem~\ref{prop:numTreeSRLG}, we may need up to $C^{|E_P|}_{k}$ spanning trees to protect arbitrary $k$ failures, which is in fact way beyond the number of spanning trees necessary to guarantee survivability. Here we would not further discuss this property in detail but leave it for future study.

\subsection{Heuristic: {SRLG-Survivable} Protecting Spanning Trees}\label{subsec:heuristic}
{Based on Theorem \ref{prop:numTreeSRLG} and Corollary \ref{pro:srlgTreeSurv}, we present an SRLG-survivable algorithm which sequentially selects protecting spanning trees and generates their corresponding mappings. We also demonstrate its extension to $k$-survivability.}

\begin{algorithm}
\KwIn{Given $G_P=(V_P, E_P)$, $G_S=(V_S, E_S)$ and SRLG set $R_E$\;}
\KwOut{Link mapping for all links in $E_S$, a protecting spanning tree set $\mathcal{T}$,  and its protected SRLG sets\;}
Initialization: set $c_e=1$ for $e\in E_P$ and $c_u=1$ for $u\in E_S$; set the protecting spanning tree set $\mathcal{T}=\emptyset$; link route $P_{u}=\emptyset, u\in E_S$, and {vector $m=(0,\cdots, 0)$ indicating} protected SRLG set with $m\in \{0,1\}^{1\times |R_E|}$\;
\For{$r \in R_E$}
{%
	\eIf{exists $\tau\in \mathcal{T}$ such that $r\cap (\cup_{u\in \tau} \hat{p_{u}})=\emptyset$}
	{\For{$u\in \tau$}{\If{$P_u$ is $\emptyset$}{$P_{u}=\hat{p}_{u}$\;}}}
	{%
		$c_e = c_e\times M,$ for all $e\in r$\;
		\For{$u\in E_S$}
		{
			\If{$P_u\neq \emptyset$}{
				$c_u = c(P_u)$\;
			}
			\Else{Generate $\hat{p_u} = \min\{p_u: p_u\in G_P\}$\;$c_u = c(\hat{p_u})$\;}
		}
		Generate the minimal spanning tree based on $c_u$: $\hat{\tau}=\min\{\sum_{u\in \tau}c_u, \tau\in G_S\}$\;
		$\mathcal{T} = \mathcal{T}\bigcup \{\hat{\tau}\}$\;
		Reset link cost: $c_e=1$ for $e\in E_P$, $c_u=1$ if $u\in \cup_{\tau\in \mathcal{T}}\tau$; otherwise, $c_u=c_u\times M$.
	}
}
\caption{{Iterative SRLG-survivable} protecting spanning tree algorithm}
\label{alg:seqTree}
\end{algorithm}
Let $M$ be {a large number}. Algorithm \ref{alg:seqTree} starts {with generating} a minimal spanning tree whose {branch} mappings are with the minimal costs.
{The algorithm then evaluates} whether existing tree(s) in the {spanning tree set} can protect at least {an} SRLG through checking whether {mappings of a tree's branches (in $\mathcal{T}$)} do not route through links in any SRLG. Then, the algorithm only creates new spanning trees for {unprotected} SRLGs. We update physical links in a unprotected SRLG set with higher costs (the physical link cost times $M$). The {newly} selected spanning tree {has} the minimal spanning tree cost ({based on} the {newly} updated physical link costs). Note here that with the new tree generation, new branch link mappings {may} be added for {the} same link in the virtual datacenter. We call it \textbf{virtual datacenter augmentation}.

Algorithm \ref{alg:seqTree} {can} easily be extended {to} $k$-survivable cloud network design {through} constructing the failure sets $R_E$ containing {failures} $r$ with $r=\{(i_1,j_1), (i_2, j_2), \cdots, (i_k,j_k)\}$, where $(i_1, j_1), (i_2,j_2),\cdots, (i_k,j_k)\in E_P$, and $(i_\ell,j_\ell)\neq (i_\beta, j_\beta)$ and $1\leq \ell,\beta \leq k$.

\section{Experiment Setting and Simulation Results}\label{sec:computation}
\begin{figure}[!ht]
\centering
\includegraphics[scale=1]{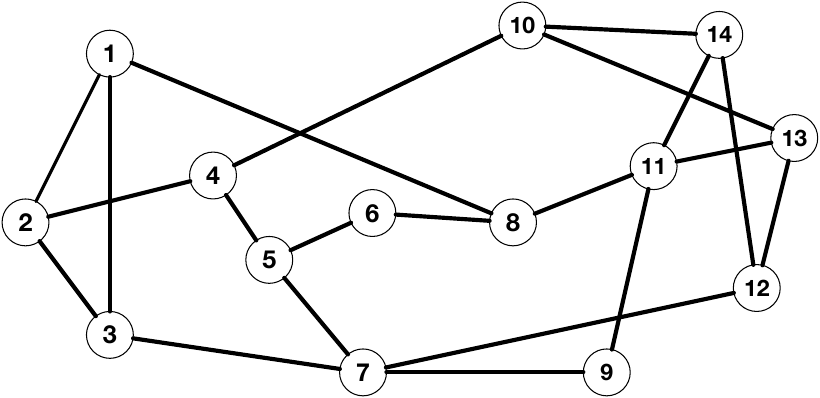}
\caption{NSF network}
\label{fig:nsf}
\end{figure}
\begin{figure}[!ht]
\includegraphics[width=0.5\textwidth]{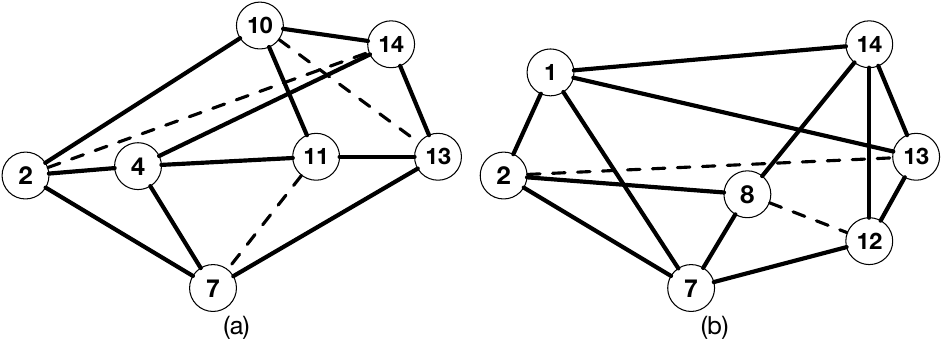}
\caption{Virtual datacenter networks}
\label{fig:cln}
\end{figure}

We choose the NSF network, illustrated in {Fig.} \ref{fig:nsf}, as the network of the physical cloud infrastructure.
{ To inspect how the nodal degree would affect the performance of our algorithms, we also create ``NSF$^{(1)}$'' which denotes the NSF network augmented with link $(6, 9)$.} For the virtual datacenter networks, we randomly generate two 3-edge connected networks denoted as ``CLN1'' and ``CLN3'' and illustrated in Fig.~\ref{fig:cln} in solid lines. { For the same rationale, we also augment ``CLN1'' and ``CLN3'' {into} 4-edge connected networks with extra links illustrated in dashed lines, which create ``CLN2'' and ``CLN4'' networks, respectively.} {The mapping of the nodes from virtual datacenter to the physical cloud infrastructure is presented through the same node indices printed in the nodes.}

Table \ref{tbl:loggraphs} provides the information of all the network topologies mentioned above, where ``Conn'', ``Nodes'', ``Edges'', ``minDeg'', ``maxDeg'', {and} ''AvgDeg'' represent the edge-connectivity, number of nodes/edges, minimal, maximal, and average node degrees of the networks.
\begin{table}[!ht]
\begin{center}
\begin{tabular}{p{0.92cm}|>{\centering\arraybackslash}m{0.5cm}
>{\centering\arraybackslash}m{0.6cm}
>{\centering\arraybackslash}m{0.6cm}
>{\centering\arraybackslash}m{0.88cm}
>{\centering\arraybackslash}m{0.88cm}
>{\centering\arraybackslash}m{0.88cm} }
\hline
\rule{0pt}{10pt}& Conn  &Nodes & Edges & MinDeg & MaxDeg & AvgDeg \\ \hline
 \rule{0pt}{8pt}CLN 1 &3  &$\;\,$7 & $\;\,$11 & {3} & {4} & 3.14 \\
 \rule{0pt}{8pt}CLN 2&4  &$\;\,$7 & $\;\,$14 & 4 & 4 & 4 \\
 \rule{0pt}{8pt}CLN 3 &3  &$\;\,$7 & $\;\,$11 & {3} & {4} & 3.29 \\
 \rule{0pt}{8pt}CLN 4&4  &$\;\,$7 & $\;\,$14 & 4 & 4 & 4 \\
 \hline
 \rule{0pt}{8pt}NSF &2 & 14 &$\;\,$21 & $\;\,$2 & 4 &  3\\
 \rule{0pt}{8pt}NSF$^{(1)}$ &3 &14 &$\;\,$22 & $\;\,$3 & 4 & 3.14 \\
\hline
\end{tabular}
\end{center}
\caption{Logical topologies information}\label{tbl:loggraphs}
\end{table}

In practice, the SRLG set is known a priori. We generate a SRLG failure set called the $3$-SRLG set, which has the following properties: (1) each SRLG has three { or fewer} edges (2) each SRLG is not a subset of another SRLG, (3) each SRLG failure does not disconnect the physical network, and (4) the union of all SRLGs covers the entire physical network~\cite{liu06}. We also consider the $k$-failure scenario where $k=2$ due to the connectivity of our physical networks.

We use two metrics {to validate and evaluate the performance of our proposed spanning-tree based approaches}: (1) full survivability and maximal partial survivability; and (2) minimal physical resources utilized to guarantee full survivability or maximal partial survivability. Note here that if a survivable mapping does not exist due to the limitation of the given networks (mainly because of their edge-connectivity), our approach would generate a mapping which guarantees the connectivity of the virtual datacenter against the most number of failure scenarios.

We report our simulation results of the proposed MILP approach in Tables~\ref{tb:3srlg_nsf1} and~\ref{tb:2plf} for the SRLG- and 2-failure scenarios, respectively. Out of the seven-element 3-SRLG set, we test our formulation over five, six, and all seven 3-SRLG failures to verify how many of the 3-SRLG failures can be protected by the generated mapping. If not all of them can be survivable, we report the maximal number of the 3-SRLGs which do not disconnect the virtual datacenter. Let ``Surv'', ``MaxS'', ``PhyS'' represent the existence of survivable cloud network mapping for the tested instances, maximal number of survivable SRLGs, and the minimal number of physical links used in the routings. Let $SIdx$ be the survivability index which shows the number of arbitrary physical link pairs whose failures do not disconnect the virtual datacenter.
\begin{table}[t]
\begin{center}
\begin{tabular}{>{\raggedright\arraybackslash}m{0.6cm}|>{\centering\arraybackslash}m{0.79cm}
|>{\centering\arraybackslash}m{0.22cm}
>{\centering\arraybackslash}m{0.34cm}
>{\centering\arraybackslash}m{0.37cm}
|>{\centering\arraybackslash}m{0.22cm}
>{\centering\arraybackslash}m{0.34cm}
>{\centering\arraybackslash}m{0.37cm}
|>{\centering\arraybackslash}m{0.22cm}
>{\centering\arraybackslash}m{0.34cm}
>{\centering\arraybackslash}m{0.37cm}
}
\hline
 \rule{0pt}{10pt} & \multirow{2}{*}{}& \multicolumn{3}{c}{Five 3-SRLGs}  &\multicolumn{3}{|c}{Six 3-SRLGs}
 &\multicolumn{3}{|c}{Seven 3-SRLGs}\\
\cline{3-11}
 \rule{0pt}{10pt} &    & Surv & MaxS & PhyS  & Surv & MaxS & PhyS & Surv & MaxS & PhyS\\
\hline
 \rule{0pt}{8pt}\multirow{4}{*}{NSF}&CLN 1&Yes&5 &22 &Yes&6 &22 &No &{6}&22\\
 \rule{0pt}{8pt}   &CLN 2&Yes&5 &27 &Yes&6 &27 &No &{6}&26\\
 \rule{0pt}{8pt}   &CLN 3&Yes&5 &26 &Yes&6 &26 &No &{6}&24\\
 \rule{0pt}{8pt}   &CLN 4&Yes&5 &31 &Yes&6 &31 &No &{6}&29\\
\hline
 \rule{0pt}{8pt}\multirow{4}{*}{$\text{NSF}^{(1)}$}&CLN 1&Yes&5 &21 &Yes&6 &21 &Yes &7&21\\
 \rule{0pt}{8pt}&CLN 2&Yes&5 &26 &Yes&6 &26 &Yes &7&26\\
 \rule{0pt}{8pt}&CLN 3&Yes&5 &23 &Yes&6 &24 &Yes &7&24\\
 \rule{0pt}{8pt}&CLN 4&Yes&5 &29 &Yes&6 &29 &Yes &7&29\\
\hline
\end{tabular}
\end{center}
\caption{MILP Results for 3-SRLG failures}
\label{tb:3srlg_nsf1}
\end{table}
\begin{table}[t]
\begin{center}
\begin{tabular}{p{1cm}|>{\centering\arraybackslash}m{0.7cm}
>{\centering\arraybackslash}m{0.85cm}
>{\centering\arraybackslash}m{0.9cm}
|>{\centering\arraybackslash}m{0.75cm}
>{\centering\arraybackslash}m{0.85cm}
>{\centering\arraybackslash}m{0.9cm}
}
\hline
 \rule{0pt}{10pt}  \multirow{2}{*}{}& \multicolumn{3}{c}{{NSF}}  &\multicolumn{3}{|c}{{NSF$^{({1})}$}}\\
\cline{2-7}
 \rule{0pt}{10pt}     & Surv & {SIdx} & PhyS  & Surv & {SIdx} & PhyS\\
\hline
 \rule{0pt}{8pt}CLN 1&No &209 &30 &No &230 &37 \\
 \rule{0pt}{8pt}CLN 2&Yes&210 &40 &Yes&231 &40 \\
 \rule{0pt}{8pt}CLN 3&Yes&210 &31 &Yes&231 &29  \\
 \rule{0pt}{8pt}CLN 4&Yes&210 &38 &Yes&231 &37  \\
\hline
\end{tabular}
\end{center}
\caption{MILP Results for 2-link failures}
\label{tb:2plf}
\end{table}

The results in Table~\ref{tb:3srlg_nsf1} show that with the same physical network, the routings of CLN2 and CLN4 consume more physical links than that of CLN1 and CLN3, which results from the more number of  virtual links and higher connectivity in CLN2 and CLN4. On the other hand, with the same virtual network, the routing consumes more physical links over NSF as the physical network than that with NSF$^{(1)}$, which corresponds to the fact that NSF network has fewer physical links and lower connectivity than NSF$^{(1)}$. At least for all testing cases, the connectivity of both virtual and physical networks impact the utilization of physical link consumption in routes of virtual links. In a cross-layer network, higher virtual network connectivity and lower physical network connectivity lead to more physical link consumption. {It is not surprising that when the physical network has lower connectivity, the survivable virtual link routings generated may require longer physical paths. Similarly, with the same physical network, the number of routings to be generated for a highly-connected virtual network is larger than those with lower connectivity.}

With Table~\ref{tb:2plf}, we observe that with the same physical network and CLN1 as virtual network, there is no survivable design for the cross-layer network. Meanwhile, CLN1 is with smallest average node degree compared with CLN2, CLN3, and CLN4. On the other hand, the node mapping between virtual network and physical network are different between CLN1-2 and CLN3-4. CLN1-2 maps onto physical node 2, 4, and 7 at the same time, the shortest paths between (2,4) and (2,7) have joint part. The mapping may lead to less physical link consumption of CLN3-4 than CLN1-2. At least, for our testing cases, the lower average virtual node degree leads to more unsurvivable cross-layer topology. Meanwhile, the node mapping between virtual network and physical network impacts the physical link consumption for virtual link routing.

As shown in Table~\ref{tb:2plf}, with NSF as the physical network, our approach can generate survivable mappings for all virtual datacenter networks against five and six 3-SRLG failures. But none of them could produce survivable cloud mappings with seven 3-SRLGs.  After augmenting NSF to NSF$^{(1)}$, all virtual datacenter networks remain connected against all five, six, and seven 3-SRLG failures, { which shows that SRLG-survivability is sensitive to the physical node degrees.} Meanwhile, it can be observed that { the utilization of physical links decreases for some tested instances when increasing the number of 3-SRLGs}. { Though it seems counterintuitive, it actually reflects that the objective of the MILP formulations helps choose shorter routings effectively while maintaining the (maximal) survivability.} All tested cases are solved within 1469 seconds.

\begin{figure}
\centering
\includegraphics[scale=0.55]{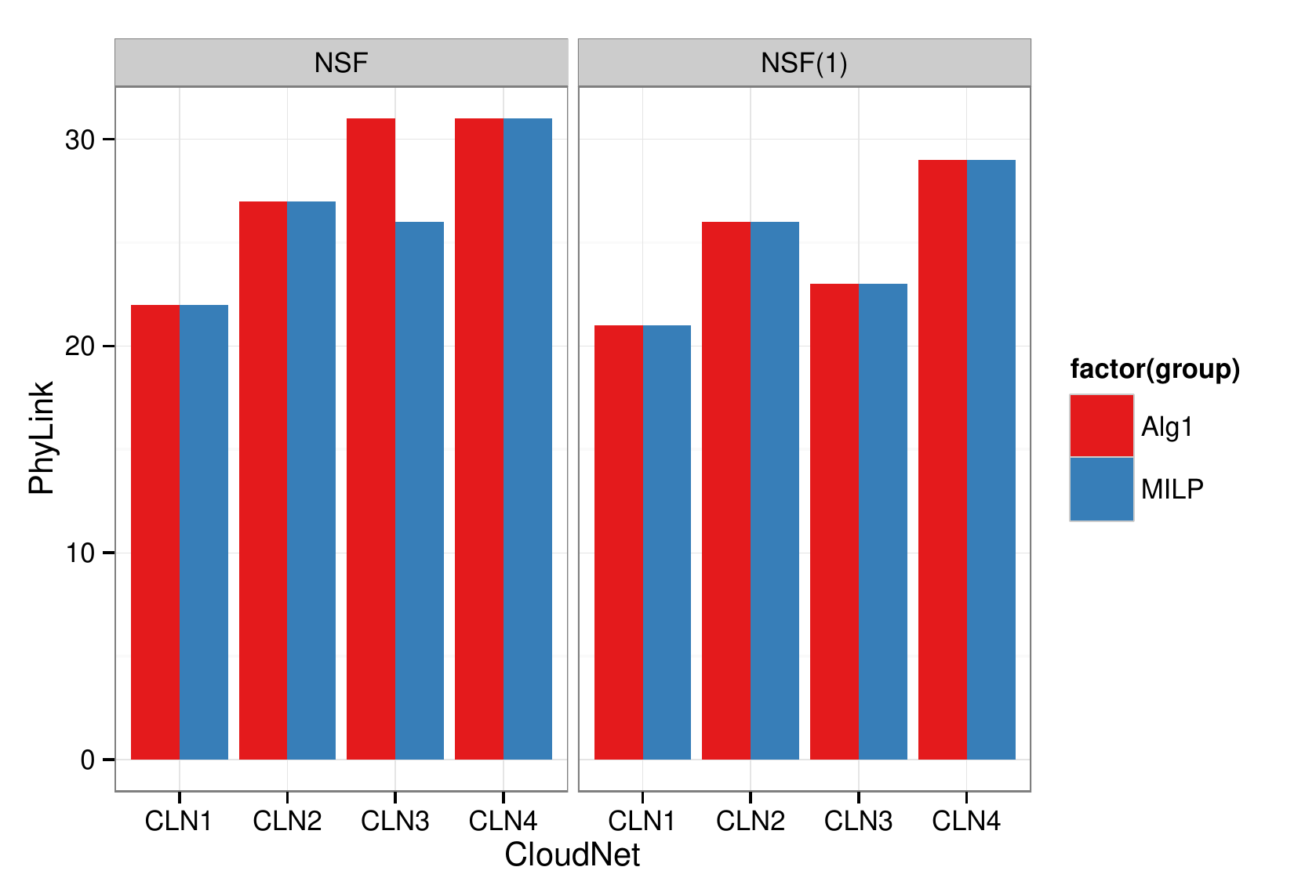}
\caption{Performance comparison between Algorithm 1 and MILP with 5 SRLG sets}
\label{fig:5srlg}
\end{figure}

\begin{figure}
\centering
\includegraphics[scale=0.55]{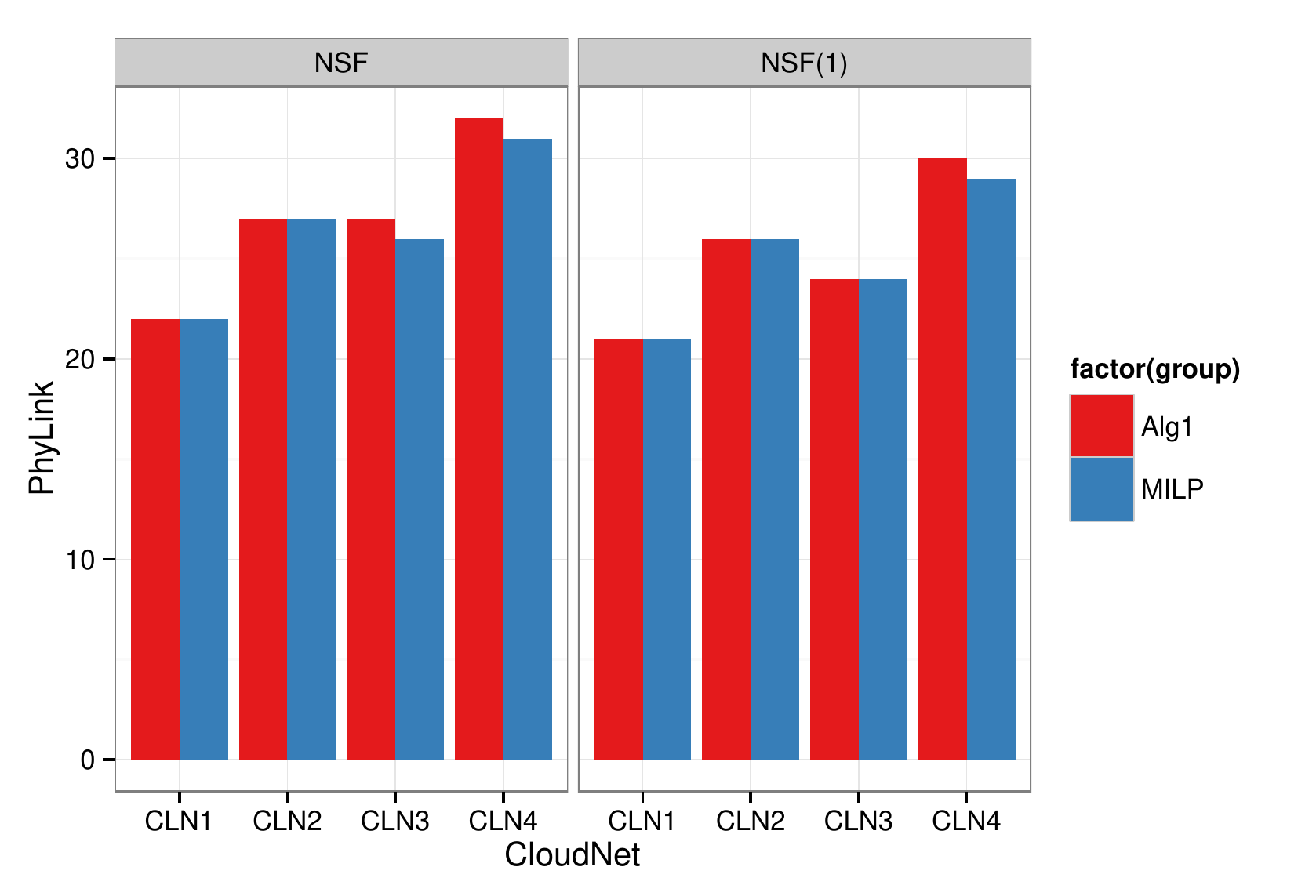}
\caption{Performance comparison between Algorithm 1 and MILP with 6 SRLG sets}
\label{fig:6srlg}
\end{figure}

\begin{figure}
\centering
\includegraphics[scale=0.55]{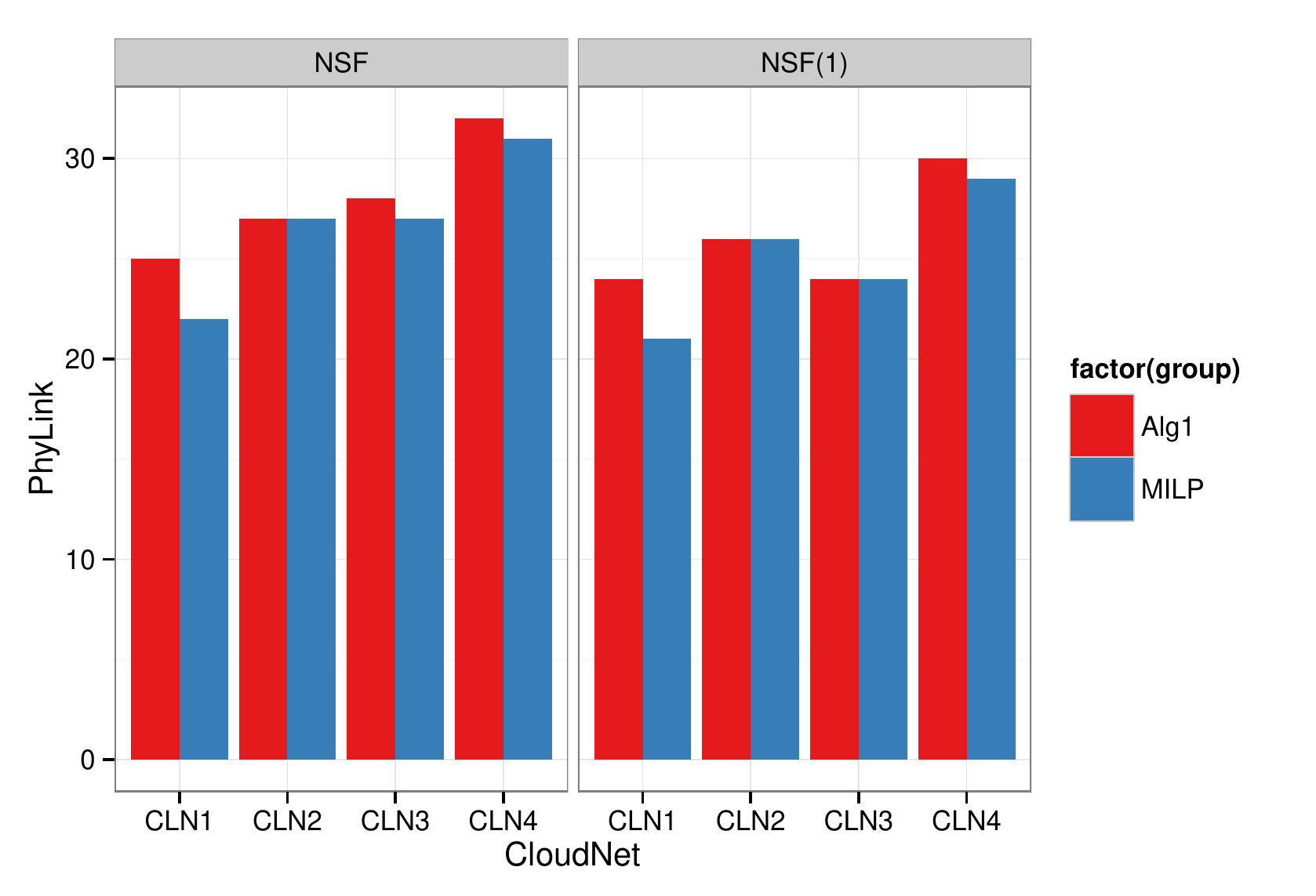}
\caption{Performance comparison between Algorithm 1 and MILP with 7 SRLG sets}
\label{fig:7srlg}
\end{figure}

We now evaluate the performance of our proposed Algorithm \ref{alg:seqTree} based on (1) whether the algorithm can provide SRLG-survivable  mapping, (2) how many SRLGs can be protected if a survivable mapping does not exist, and (3) whether cloud network link augmentation should be applied to provide better protection. The results are presented in Tables \ref{tbl:heuri2F}, \ref{tbl:heuri_5}, \ref{tbl:heuri_6}, and \ref{tbl:heuri_7}, where ``Augment\#'', ``Tree\#'', ``LogS'' and ``Time'' denote the number of augmented logical links to produce a survivable cloud mapping, the number of spanning trees generated in the virtual datacenter, the number of links utilized in the generated spanning tree set, and the computational time (in second), respectively. We illustrate the number of physical
link utilization by MILP and heuristic approaches for 5, 6, and 7 SRLG sets in Figs.~\ref{fig:5srlg}-\ref{fig:7srlg}.

\begin{table}
\begin{center}
\begin{tabular}{p{1cm}|>{\centering\arraybackslash}m{0.75cm}
>{\centering\arraybackslash}m{0.85cm}
>{\centering\arraybackslash}m{0.9cm}
|>{\centering\arraybackslash}m{0.75cm}
>{\centering\arraybackslash}m{0.85cm}
>{\centering\arraybackslash}m{0.9cm}
}
\hline
\rule{0pt}{10pt} &\multicolumn{3}{c|}{NSF} & \multicolumn{3}{c}{NSF$^{(1)}$}\\
\cline{2-7}
\rule{0pt}{10pt} &PrctFSet &PrctPct &Time(s) &PrctFSet &PrctPct &Time(s)\\
\hline
\rule{0pt}{8pt} CLN1& -/210 &-&-&-/231&-&-\\
\rule{0pt}{8pt} CLN2&200/210&95.24\%&13.95&221/231&95.67\%&13.86\\
\rule{0pt}{8pt} CLN3&182/210&86.67\%&12.98&202/231&87.01\%&13.51\\
\rule{0pt}{8pt} CLN4&209/210&99.52\%&4.45 &228/231&98.70\%&5.83\\
\hline
\end{tabular}
\caption{Algorithm 1  results for 2-link failure}
\label{tbl:heuri2F}
\end{center}
\end{table}

\begin{table}[t]
\centering
\begin{tabular}{>{\raggedright\arraybackslash}m{0.5cm}
>{\centering\arraybackslash}m{0.4cm}
>{\centering\arraybackslash}m{0.3cm}
>{\centering\arraybackslash}m{1.1cm}
>{\centering\arraybackslash}m{0.3cm}
>{\centering\arraybackslash}m{0.3cm}
>{\centering\arraybackslash}m{0.55cm}
>{\centering\arraybackslash}m{0.5cm}
>{\centering\arraybackslash}m{0.65cm}
}
\hline
 \rule{0pt}{8pt}   & &Surv       &Augment\# & MaxS & PhyS & Tree\# & LogS & Time(s)\\
\hline
 \rule{0pt}{8pt}\multirow{4}{*}{NSF}&CLN1&Yes     &0  &5  &22 &4 &10 &0.078\\
 \rule{0pt}{8pt}   &CLN2&Yes     &0  &5  &27 &4 &9 &0.140\\
 \rule{0pt}{8pt}   &CLN3&Yes     &0  &5  &27 &5 &11 &0.140\\
 \rule{0pt}{8pt}   &CLN4&Yes     &0  &5  &31 &5 &11 &0.062\\
\hline\hline
 \rule{0pt}{8pt}\multirow{4}{*}{$\text{NSF}^{(1)}$}&CLN1&Yes     &0  &5  &21 &5 &10 &0.124\\
 \rule{0pt}{8pt}&CLN2&Yes     &0  &5  &26 &4 &9 &0.078\\
 \rule{0pt}{8pt}&CLN3&Yes     &0  &5  &23 &5 &11 &0.102\\
 \rule{0pt}{8pt}&CLN4&Yes     &0  &5  &29 &5 &10 &0.156\\
\hline
\end{tabular}
\caption{Algorithm \ref{alg:seqTree} results with five 3-SRLG sets}
\label{tbl:heuri_5}
\end{table}

{We present the performance of Algorithm 1 for 2-link failure scenario in Table~\ref{tbl:heuri2F}. With NSF and NSF$^{(1)}$ as the physical networks, the total number of 2-link failures is 210 and 231, respectively. Let ``PrctFSet'' represent the number of protected failure set and ``PrctPct'' as the percentage of the number of protected sets to the total number of failure sets. The two numbers in column ``PrctFSet'' represent the number of protected set followed by the total number of failure sets. The best performance achieved by our algorithm is with NSF and CLN4 as the physical and virtual networks. The average protected percentage for all these testing cases are 93.81\%.}

{There is no surprise that the MILP approach's computation time is much longer than that of the heuristic algorithm. The average computation time with our MILP approach for survivable routing with 3-SRLG failure and 2-failure are 11.93 and 194.88 seconds, respectively. Among all testing cases, the longest computation time is 852.01 seconds (2-failure, with NSF and CLN2 as the physical and logical networks, respectively). As shown in Tables VI-VIII, the computation time for our heuristic algorithm over all test cases is less than 0.16 seconds.}

It can be observed in Tables~\ref{tbl:heuri_5} to~\ref{tbl:heuri_7} that: (1) if SRLG-survivable cloud network mapping exists, the results produced by Algorithm \ref{alg:seqTree} are as good as the ones generated by the MILP formulation, (2) the number physical links utilized in the routings produced by Algorithm \ref{alg:seqTree} is the same or close to the results of MILP formulation, and (3) for a physical network with higher node degree, the generated cloud network mappings seem to utilize less number of links both in the physical cloud infrastructure and the virtual datacenter to construct all protecting spanning trees and their routings.

{ As expected, Algorithm \ref{alg:seqTree} cannot generate the survivable cloud network mappings for all instances in Table~\ref{tbl:heuri_7}, which triggers the augmentation of link(s) in the virtual datacenter network. We wish to note here that we only augment CLN1 with both NSF and NSF$^{(1)}$as the physical networks because (1) we want to do both vertical (between NSF and  NSF$^{(1)}$) and horizontal (between Algorithm and MILP) comparisons, and (2) we want to show how many 3-SRLGs can be protected without augmentation. As seen in Table~\ref{tbl:heuri_7}, though not all mappings generated by the heuristic are survivable, they still can protect majority number of 3-SRLG failures.}
\begin{table}[t]
\centering
\begin{tabular}{>{\raggedright\arraybackslash}m{0.6cm}
>{\centering\arraybackslash}m{0.4cm}
>{\centering\arraybackslash}m{0.3cm}
>{\centering\arraybackslash}m{1.1cm}
>{\centering\arraybackslash}m{0.3cm}
>{\centering\arraybackslash}m{0.3cm}
>{\centering\arraybackslash}m{0.55cm}
>{\centering\arraybackslash}m{0.5cm}
>{\centering\arraybackslash}m{0.65cm}
}
\hline
 \rule{0pt}{8pt}   & &Surv        &Augment\# & MaxS & PhyS & Tree\# & LogS & Time(s)\\
\hline
 \rule{0pt}{8pt}\multirow{4}{*}{NSF}&CLN1&Yes    &0  &6  &22 &5 &10 &0.100\\
 \rule{0pt}{8pt}&CLN2&Yes     &0  &6  &27 &5 &10 &0.120\\
 \rule{0pt}{8pt}&CLN3&Yes     &0  &6  &27 &6 &11 &0.083\\
 \rule{0pt}{8pt}&CLN4&Yes     &0  &6  &32 &6 &11 &0.112\\
 \hline\hline
 \rule{0pt}{8pt}\multirow{4}{*}{$\text{NSF}^{(1)}$}&CLN1&Yes     &0  &6  &21 &5 &10 &0.124\\
 \rule{0pt}{8pt}&CLN2&Yes     &0  &6  &26 &5 &9 &0.078\\
 \rule{0pt}{8pt}&CLN3&Yes     &0  &6  &24 &5 &11 &0.096\\
 \rule{0pt}{8pt}&CLN4&Yes     &0  &6  &30 &5 &11 &0.113\\
 \hline
\end{tabular}
\caption{Algorithm \ref{alg:seqTree} results with six 3-SRLG sets}
\label{tbl:heuri_6}
\end{table}
\begin{table}[t]
\centering
\begin{tabular}{>{\raggedright\arraybackslash}m{0.6cm}
>{\centering\arraybackslash}m{0.4cm}
>{\centering\arraybackslash}m{0.3cm}
>{\centering\arraybackslash}m{1.1cm}
>{\centering\arraybackslash}m{0.3cm}
>{\centering\arraybackslash}m{0.3cm}
>{\centering\arraybackslash}m{0.55cm}
>{\centering\arraybackslash}m{0.5cm}
>{\centering\arraybackslash}m{0.65cm}
}
\hline
 \rule{0pt}{8pt}   & &Surv        & Augment\# & MaxS & PhyS & Tree\# & LogS & Time(s)\\
\hline
 \rule{0pt}{8pt}\multirow{4}{*}{NSF}&CLN1&No    &1  &5  &25 &5 &10 &0.085\\
 \rule{0pt}{8pt}&CLN2&No     &0  &6  &27 &6 &10 &0.093\\
 \rule{0pt}{8pt}&CLN3&No     &0  &6  &28 &6 &12 &0.110\\
 \rule{0pt}{8pt}&CLN4&No     &0  &6  &32 &6 &13 &0.123\\
 \hline\hline
 \rule{0pt}{8pt}\multirow{4}{*}{$\text{NSF}^{(1)}$}&CLN1&No     &1  &6  &24 &6 &10 &0.081\\
 \rule{0pt}{8pt}&CLN2&Yes     &0  &7  &26 &6 &10 &0.097\\
 \rule{0pt}{8pt}&CLN3&Yes    &0  &7  &24 &5 &11 &0.116\\
 \rule{0pt}{8pt}&CLN4&Yes     &0  &7  &30 &5 &11 &0.146\\
\hline
\end{tabular}
\caption{Algorithm \ref{alg:seqTree} results with seven 3-SRLG sets}
\label{tbl:heuri_7}
\end{table}


\section{Conclusion}\label{sec:conclusion}
In this paper, we studied the SRLG- and $k$-survivable cloud network mapping problems and discussed how the concept of the spanning tree can be used to guarantee survivability. Based on the spanning tree concept, we proposed the necessary and sufficient conditions and the corresponding MILP formulations, which are applied in this paper as the general framework to handle SRLG and $k$ failures in a cloud network. In addition to the mathematical formulation, we designed a heuristic algorithm also based on the spanning tree concept. We compared the performance of the simulation results generated by both the MILP formulation and heuristic algorithm, which showed that our proposed Algorithm \ref{alg:seqTree} can effectively generate SRLG-survivable cloud network mappings with quality similar to the optimal solution generated by the MILP formulation. Also, the difference of physical resource utilization between Algorithm \ref{alg:seqTree} and the optimal solution is within 5\% in average {and within 17\%} for all tested cases.

\section*{Appendix}
This appendix includes three parts. We first present the cutset-based  cloud network mapping formulation for $k$- and SRLG-survivability, which is the counterpart of our  spanning-tree based approach. We also provide their corresponding MILP formulations.  The second part provides the comparisons of complexity for the cutset- and spanning-tree-based approaches in terms of the number of formulations generated. The last one provides the proofs for Lemmas~\ref{lemma:2FailureMset} and~\ref{lemma:kFailureMset}.

\subsection{Cutset-Based Necessary and Sufficient Conditions for Survivable Cloud Network Mapping}\label{subsec:cutset}
Let $\chi$ {be} a cutset in the virtual datacenter network, $\chi \subset G_S$.

\begin{lemma}
A cloud network {mapping} is 1-survivable if and only if for each $(i,j)\in E_P$,
\begin{align}
|\chi \cap \Lambda(i,j)| \leq |\chi|-1, \;\;\;\; \text{for all cutsets $\chi \subset G_S$,}\nonumber
\end{align}
\end{lemma}
where $|\cdot|$ represents the cardinality of the set. This lemma follows {the} fact that a mapping is survivable if and only if {any $(i,j)$ failure would not disconnect all links in every single cutset in the virtual datacenter}.

Extending this {lemma}, we get the following necessary and sufficient conditions for $k$-survivability.
\begin{theorem}\label{thm:cutset_ksurv}
A cloud network mapping is $k$-survivable if and only if {after} arbitrary $k$ physical link {failures},
\begin{align}
\left|\chi \bigcap \left(\bigcup_{\beta=1}^{k}\Lambda(i_\beta, j_\beta)\right)\right| \leq |\chi|-1,\;\;\;\text{$\forall \chi \subset G_S$,}\label{cutIneq_k}
\end{align}
where $(i_1, j_1)\neq (i_2, j_2)\neq \cdots\neq (i_k, j_k)$ and $(i_1, j_1), (i_2,j_2), \cdots, (i_k,j_k)\in E_P$.
\end{theorem}
\begin{IEEEproof}
{Necessary condition: after any {$k$-link failures}, if all cuts are not fully disconnected (at least an edge in the cut remains connected), the virtual datacenter remains connected. Hence, the cloud network mapping is $k-$survivable. \\
Sufficient condition: If a cloud network is $k-$survivable, the virtual datacenter is connected after any $k$-link failures, hence, no cut is fully disconnected.}
\end{IEEEproof}

\begin{corollary}\label{co:cutset_treeTheorem}
A cloud network mapping is SRLG-survivable if and only if after any $r\in R_E$ failure,
\begin{align}
\left|\chi \bigcap \left(\bigcup_{(i^{r}_{k(r)}, j^{r}_{k(r)})\in r}\Lambda(i^{r}_{k(r)},j^{r}_{k(r)}) \right)\right|\leq |\chi|-1,\nonumber\\\;\;\;\text{for all cutsets $\chi \subset G_S$.}\label{cutIneq_srlg}
\end{align}
\end{corollary}
The corollary above is a direct extension of Theorem \ref{thm:cutset_ksurv}.

Based on Theorem \ref{thm:cutset_ksurv} and Lemma~\ref{lemma:kFailureMset}, we provide another MILP formulation which guarantees that the cloud network mapping is $k$-survivable. Following  Lemma \ref{lemma:kFailureMset}, the formulation determines the relationship between $y^{st}_{i_1 j_1,\ldots,i_k j_k}$ and $y^{st}_{i_\beta j_\beta}$, $1\leq \beta\leq k$, which indicates whether $(s,t)$ is routed through links in a subset of $\{(i_1,j_1), \cdots, (i_k,j_k)\}$.

\begin{proposition}\label{prop:cutsetEvalute}
A cloud network mapping is $k$-survivable, $k\geq 1$, if and only if the following condition holds:
\begin{align}
&\sum_{(s,t)\in \chi}y^{st}_{i_1j_1,\cdots,i_k,j_k} \leq |\chi|-1 ,\nonumber\\
&\text{where }\chi\subset G_S,\;\; (i_1,j_1), \cdots, (i_k,j_k)\in E_P,\nonumber\\
&\text{ and } (i_q,j_q)\neq (i_\ell, j_\ell),\;\; 1\leq q, \ell \leq k,\label{form:cutset_k_surv}\\
&y^{st}_{i_1j_1,\cdots,i_k,j_k} \in \{0,1\}. \label{eq:cut_binary}
\end{align}
\end{proposition}
It is easy to verify that $\sum_{(s,t)\in \chi}y^{st}_{i_1j_1,\cdots,i_k,j_k}= |\chi \bigcap \bigcup^{k}_{\beta-1}\Lambda(i_\beta,j_\beta)|$ based on the definition of $y^{st}_{i_1j_1,\cdots,i_k,j_k}$, when $\{(i_1,j_1),\cdots, (i_k,j_k)\}$ is a $k$-link failure.

We provide the ILP formulation for $k-$survivability based on the cutset concept as follows:
\begin{align}
&\min\sum_{(s,t)\in E_S}\sum_{(i,j)\in E_P}y^{ij}_{st}\nonumber\\
s.t.\;&\text{Constraint (\ref{fm:2FailMseta}) to (\ref{fm:2FailMsetc}),  (\ref{eq:lightpath1}), (\ref{eq:feasible_y}), (\ref{form:cutset_k_surv}) \text{and} (\ref{eq:cut_binary}).} \nonumber
\end{align}

\begin{proposition}\label{prop:cutset_srlg}
A cloud network mapping is survivable after SRLG failures if and only if the following conditions hold: for any $r\in R_E$,
\begin{align}
&\text{constraint (\ref{form:cutset_k_surv})},\nonumber\\
&y_{i^{r}_{1}j^{r}_{1}, i^{r}_{2}j^{r}_{2}, \ldots, i^{r}_{k(r)}j^{r}_{k(r)}}^{st} \geq (y^{st}_{i^{r}_{1}j^{r}_{1}} +y^{st}_{j^{r}_{1}i^{r}_{1}}), \label{eq:cut_ksurvReCapa11} \\
&y_{i^{r}_{1}j^{r}_{1}, i^{r}_{2}j^{r}_{2}, \ldots, i^{r}_{k(r)}j^{r}_{k(r)}}^{st} \geq (y^{st}_{i^{r}_{2}j^{r}_{2}} +y^{st}_{j^{r}_{2}i^{r}_{2}}), \label{eq:cut_ksurvReCapa21} \\
&\ldots\ldots \nonumber \\
&y_{i^{r}_{1}j^{r}_{1}, i^{r}_{2}j^{r}_{2}, \ldots, i^{r}_{k(r)}j^{r}_{k(r)}}^{st} \geq (y^{st}_{i^{r}_{k(r)}j^{r}_{k(r)}} +
y^{st}_{j^{r}_{k(r)}i^{r}_{k(r)}}), \label{eq:cut_ksurvReCapa31}\\
& y_{i^{r}_1j^{r}_1, i^{r}_2j^{r}_2, \ldots, i^{r}_{k(r)}j^{r}_{k(r)}}^{st} \in \{0,1\}, \quad\quad (s,t)\in E_{S}. \label{eq:cut_rrange}
\end{align}
\end{proposition}

The ILP formulation for SRLG-survivability based on the concept of cutset is as follows:
\begin{align}
&\min\sum_{(s,t)\in E_S}\sum_{(i,j)\in E_P}y^{ij}_{st}\nonumber\\
s.t.\;&\text{Constraint (\ref{fm:2FailMseta}) to (\ref{fm:2FailMsetc}), (\ref{eq:lightpath1}), (\ref{eq:feasible_y}), and (\ref{form:cutset_k_surv}) to (\ref{eq:cut_rrange}).} \nonumber
\end{align}

\subsection{Complexity Comparision of the Spanning-Tree and Cutset Based Approaches}
The complexity of MILP models for both spanning-tree and cutset-based formulations in terms of the numbers of variables and constraints for SRLG- and $k$-survivability is shown in Table~\ref{tb:complexity}. Let ``Tree'' and ``Cutset'' represent the spanning-tree and cutset based MILP formulations, respectively. Here $|V_P|$, $|E_P|$, $|V_S|$, $|E_S|$ represent the node and edge numbers in the physical cloud infrastructure and virtual datacenter, respectively. We let $r_{\max}$ be an SRLG set with the maximum number of physical edges in $R_E$ and $N$ represent the number of cutsets in $G_S$.
\begin{table}[!ht]
\begin{center}
\begin{tabular}{p{0.4cm}|>{\centering\arraybackslash}m{0.6cm}|>{\centering\arraybackslash}m{3.2cm}
|>{\centering\arraybackslash}m{3.2cm}}
\hline
  \rule{0pt}{8pt}   &Model & SRLG & $k$-survivable\\
\hline
 \rule{0pt}{11pt} \multirow{2}{*}{Vrbs}& Tree  & $\mathcal{O}[|E_S|(|R_E|+|E_P|)]$
 & $\mathcal{O}[|E_S|(C_{k}^{|E_P|}+|E_P|)]$\\
\cline{2-4}
 \rule{0pt}{14pt}    & Cutset  &  $\mathcal{O}[|E_S|(|R_E|+|E_P|)]$
 & $\mathcal{O}[|E_S|(C_{k}^{|E_P|}+|E_P|)]$\\
\hline
 \rule{0pt}{14pt} \multirow{2}{*}{Cnts} &Tree & $\mathcal{O}[|E_S||V_P|+|R_E|(|E_S|+|V_S|)]$
  & $\mathcal{O}[|E_S||V_P|+ C_{k}^{|E_P|}(|E_S|+ |V_S|)]$ \\
 \cline{2-4}
  \rule{0pt}{14pt}  &Cutset & $\mathcal{O}[|E_S||V_P|+|R_E|(|E_S|+|V_S|+2^{|V_S|})]$
   & $\mathcal{O}[|E_S||V_P|+C_{k}^{|E_P|}(|E_S|+|V_S|+2^{|V_S|})]$\\
\hline
\end{tabular}
\end{center}
\caption{Complexity of MILP formulations with number of variables and constraints}
\label{tb:complexity}
\end{table}

We observe that the complexity of the formulation depends mainly on the number of failure scenarios. Given an SRLG set, the complexity of the formulation is bounded by the cardinality of the SRLG set, the number of physical links in an SRLG set, and the size of the physical infrastructure and virtual datacenter. But for the generalized $k$-failure problem, the complexity of the formulation increases exponentially with the number of failed links. {Comparing the complexity of tree-based and cutset-based MILP formulations, the number of variables are the same for both formulations, but the tree-based formulation uses much less binary variables. The number of constraints in the cutset-based model is significantly more than that based on spanning trees because the cutset-based formulation needs to enumerate all cutsets in the physical network, which significantly increases the number of constraints. In general, it takes longer time to compute the optimal result for an MILP formulation with more binary variables and constraints. Hence, in this paper our focus is mainly on the MILP formulation and heuristic based on the spanning tree concept.}

\subsection{Proofs}\label{appendix:proofs}
\textbf{Proof of Lemma~\ref{lemma:2FailureMset}:}
\begin{IEEEproof}
Necessary condition:
given a $(s,t)\in E_S$ and two physical {links} $(i_1, j_1), (i_2, j_2)\in E_P$,
{one of} the following four cases {occurs}:
(i) $(s,t)\in \Lambda(i_1, j_1)\cap \overline{\Lambda}(i_2, j_2)$;
(ii) $(s,t)\in \overline{\Lambda}(i_1, j_1)\cap \Lambda(i_2, j_2);
(iii) (s,t)\in \Lambda(i_1, j_1)\cap \Lambda(i_2, j_2)$;
or (iv) $(s,t) \in \overline{\Lambda}(i_1, j_1)\cap\overline{\Lambda}(i_2, j_2)$.
For case (i), $y^{st}_{i_1j_1}+y^{st}_{j_1i_1}=1$, so $y^{st}_{i_1j_1, i_2j_2}=1$;
for case (ii), $y^{st}_{i_2j_2}+y^{st}_{j_2i_2}=1$, then, $y^{st}_{i_1j_1, i_2j_2}=1$;
for case (iii), $y^{st}_{i_1j_1}+y^{st}_{j_1i_1}=1$ and {$y^{st}_{i_2j_2}+y^{st}_{j_2i_2}=1$}, then, $y^{st}_{i_1j_1, i_2j_2}=1$;
for case (iv), $y^{st}_{i_1j_1}=y^{st}_{j_1i_1}=0$ and $y^{st}_{i_2j_2}=y^{st}_{j_2i_2}=0$, then, $y^{st}_{i_1j_1, i_2j_2}=0$.
In all cases, constraints (\ref{fm:2FailMseta}) to (\ref{fm:2FailMsetc}) hold.

Sufficient condition: with constraints (\ref{fm:2FailMseta})--(\ref{fm:2FailMsetc}), if either $y^{st}_{i_1j_1}+y^{st}_{j_1i_1}=1$ or $y^{st}_{i_2j_2}+y^{st}_{j_2i_2}=1$, then, $y^{st}_{i_1j_1, i_2j_2}=1$, which implies that $(s,t)\in \Lambda(i_1, j_1)\cup \Lambda(i_2,j_2)$.
Meanwhile, if $y^{st}_{i_1j_1}=y^{st}_{j_1i_1}=y^{st}_{i_2j_2}=y^{st}_{j_2i_2}=0$, then, $y^{st}_{i_1j_1, i_2j_2}=0$, which implies that $(s,t)\in \overline{\Lambda}(i_1, j_1)\cap\overline{\Lambda}(i_2, j_2)$.
\end{IEEEproof}

\textbf{Proof of Lemma~\ref{lemma:kFailureMset}:}
\begin{IEEEproof}We prove this conclusion by induction for both necessary and sufficient conditions. \\
Necessary condition: with Lemma \ref{lemma:2FailureMset}, if $\beta =2$, the conclusion holds. We assume that if $\beta=k-1$, the conclusion holds. With $\beta=k$, four cases {occur} with $[(i_1,j_1), \cdots, (i_{k-1},j_{k-1})]$ and $(i_k, j_k)$ failures: (i) {$(s,t)$'s} mapping routes through $(i_k, j_k)$ only; (ii) {$(s,t)$'s} mapping routes through {all or} part of $[(i_1,j_1), \cdots, (i_{k-1},j_{k-1})]$; (iii) {$(s,t)$'s} mapping routes through {all or} part of $[(i_1,j_1), \cdots, (i_{k-1},j_{k-1})]$ and $(i_k, j_k)$; and (iv) {$(s,t)$'s}  mapping {routes} though none of $\{(i_{{q}}, j_{{q}})\text{ with } 1\leq {q} \leq k \}$. For case (i), $y^{st}_{i_kj_k}=1$, {so} $y^{st}_{i_1j_1, \cdots, i_kj_k}=1$. For case (ii), with ${q}=k-1$, there exists {some} ${z}$ (${1} \leq {z}\leq k-1$), such that {$y^{st}_{i_zj_z}+y^{st}_{j_zi_z}=1$}, {so} $y^{st}_{i_1j_1, \cdots, i_{k-1}j_{k-1}}=1$. Hence, $y^{st}_{i_1j_1, \cdots, i_{k}j_{k}}=1$. For case (iii), if both $y^{st}_{i_kj_k}=1$ and $y^{st}_{i_1j_1,\cdots,i_{k-1}j_{k-1}}=1$, $y^{st}_{i_1j_1,\cdots,i_{k}j_{k}}=1$. These three cases imply that constraints (\ref{beta_fail_1})--(\ref{beta_fail_beta}) hold. For case (iv), if $y^{st}_{i_1j_1}=y^{st}_{j_1i_1}=\cdots= y^{st}_{i_kj_k}=y^{st}_{j_ki_k}=0$, then $y^{st}_{i_1j_1,\cdots, y_kj_k}=0$, which leads to constraint (\ref{beta_fail_all}).\\
Sufficient condition: if $\beta=2$, {the conclusion holds with {Lemma 3}.}
{We assume that the conclusion holds when $\beta=k-1$.}
{Now we prove that {the} conclusion holds when $\beta=k$. {Here,} only two cases {may} {occur}: (i)$(s,t)\in \Lambda(i_k, j_k)$, and (ii) $(s,t)\in \bar{\Lambda}(i_k, j_k)$.}
{For case (i), with $y^{st}_{i_1j_1,\cdots, i_kj_k}\geq y^{st}_{i_kj_k} {+ y^{st}_{j_ki_k}}$}, {it implies $(s,t)\in \cup_{\kappa=1}^{k}\Lambda(i_\kappa,j_\kappa)$}. {For case (ii),
with $\beta=k-1$, if constraints (\ref{beta_fail_1})--(\ref{beta_fail_beta}) lead to $(s,t)\in \cup_{\kappa=1}^{k-1}\Lambda(i_\kappa,j_\kappa)$, then $(s,t)\in \cup_{q=1}^{k}\Lambda(i_q, j_q)$; otherwise, $y^{st}_{i_1j_1,\cdots, i_{k-1}j_{k-1}}\leq \sum_{q=1}^{k-1}{y^{st}_{i_q j_q + y^{st}_{j_q i_q}}}$ which implies $(s,t)\in \cap_{q=1}^{k-1}\overline{\Lambda}(i_q,j_q)$. With $y^{st}_{i_1j_1,\cdots, i_{k-1}j_{k-1}}\leq \sum_{q=1}^{k-1}y^{st}_{i_q j_q} {+ y^{st}_{j_q i_q}} + y_{i_kj_k} {+ y_{j_ki_k}}$, it implies $(s,t)\in \cap_{q=1}^{k}\overline{\Lambda}(i_q,j_q)$.} Hence, the conclusion holds when $\beta=k$.
\end{IEEEproof}

\bibliographystyle{IEEEtran}
\bibliography{p3}

\end{document}